# Imaging transformer for MRI denoising with the SNR unit training: enabling generalization across field-strengths, imaging contrasts, and anatomy

Hui Xue[1], Sarah Hooper[1], Azaan Rehman[1], Iain Pierce[2], Thomas Treibel[2], Rhodri Davies[2], W Patricia Bandettini[1], Rajiv Ramasawmy[1], Ahsan Javed[1], Zheren Zhu[3], Yang Yang[3], James Moon[2], Adrienne Campbell[1], and Peter Kellman[1]

[1]National Heart, Lung, and Blood Institute, Bethesda, MD, United States
[2]Barts Heart Centre at St. Bartholomew's Hospital, London, United Kingdom
[3]University of California, San Francisco, CA, United States

## Corresponding author:

Hui Xue

National Heart, Lung and Blood Institute

National Institutes of Health

10 Center Drive, Bethesda

MD 20892

USA

Phone:  +1 (301) 827-0156

Cell:      +1 (609) 712-3398

Fax:      +1 (301) 496-2389

Email: hui.xue@nih.gov

**Word Count: 1,797**

**Abbreviations**

SNR = signal noise ratio

DL = deep learning

EF = ejection fraction

LV = left ventricular

ESV = end-systolic volume

EDV = end-diastolic volume

GPU = graphical processing units

CMR = cardiac magnetic resonance

CNN = convolutional neural networks

imformer = imaging transformer

**Abstract**

The ability to recover MRI signal from noise is key to achieve fast acquisition, accurate quantification, and high image quality. Past work has shown convolutional neural networks can be used with abundant and paired low and high-SNR images for training. However, for applications where high-SNR data is difficult to produce at scale (e.g. with aggressive acceleration, high resolution, or low field strength), training a new denoising network using a large quantity of high-SNR images can be infeasible.

In this study, we overcome this limitation by improving the generalization of denoising models, enabling application to many settings beyond what appears in the training data. Specifically, we **a)** develop a training scheme that uses complex MRIs reconstructed in the SNR units (i.e., the images have a fixed noise level, **SNR unit training**) and augments images with realistic noise based on coil g-factor, and **b)** develop a novel **imaging transformer (imformer)** to handle 2D, 2D+T, and 3D MRIs in one model architecture. Through empirical evaluation, we show this combination improves performance compared to CNN models and improves generalization, enabling a denoising model to be used across field-strengths, image contrasts, and anatomy.

Introduction

The ability to recover MRI signal from noise is key to achieve fast acquisition, accurate quantification, and high image quality. Past work has shown convolutional neural networks (CNNs) [1] can be used with abundant and paired low and high-SNR images for training. However, for applications where high-SNR data is difficult to produce at scale (e.g. with aggressive acceleration, high resolution, or low field strength), training a new denoising network using a large quantity of high-SNR images can be infeasible. In this study, we overcome this limitation by improving the generalization of denoising models, enabling application to many settings beyond what appears in the training data. Specifically, we **a)** develop a training scheme that uses complex MRIs reconstructed in the SNR units (i.e., the images have a fixed noise level) and augments images with realistic noise based on coil g-factor, and **b)** develop a novel imaging transformer (imformer) to handle 2D, 2D+T, and 3D MRIs in one model architecture. Through empirical evaluation, we show this combination improves performance compared to CNN models and improves generalization, enabling a denoising model to be used across field-strengths, image contrasts, and anatomy.

Methods

The proposed training method is shown in Fig. 1. High SNR complex images (cardiac cine acquired solely in 3T scanners) were degraded with realistic MR noise generated by applying g-factor maps (computed at R=2 to 6), partial Fourier and kspace filters. Unlike past approaches, which typically normalize the signal levels (e.g., to range [0, 1]), we focus on normalized the noise level. The SNR unit [2] reconstruction was applied throughout training to keep the noise standard deviation at 1.0.

The corrupted images and g-factor maps were input into the model, with patch training and a long-term skip connection. The model was optimized to recover high-quality images from degraded inputs.

Additionally, we propose a novel imaging transformer block, shown in Fig. 2, to enable the model to process 2D, 2D+T, and 3D images. The key innovation is to incorporate three configurable imaging attention modules: spatial local (L), spatial global (G) and temporal (T) attention. A block allows any combination of many attention modules. By stacking many blocks together, different imaging transformer models can be constructed, enabling every output pixel to be computed by nonlinearly combining all input pixels with data-specific attention coefficients. Two imaging transformer architectures were implemented, inspired by two popular CNNs: Unet [3] and high-res net ("HRnet") [4]. The imaging transformer Unet and imaging transformer HRnet are shown in Fig. 2b and c, respectively.

To evaluate the proposed training scheme and imaging transformers, we ran the following experiments.

*Model comparison*. First, the imaging transformer was compared against CNNs and vision transformers. Five models were compared: three "standard" models – CNN Unet, CNN HRnet and swin3D transformer [5]; two imaging transformer models – Unet-imformer and HRnet-imformer. All models were trained on 309,238 3T retro-cine complex 2D+T series (MAGNETOM Prisma, Siemens Healthcare). The dataset was split into 90% for training and 10% for validation. In all cases, training used SNR unit reconstruction and noise augmentation. To illustrate the added value of noise augmentation, the four models were additionally trained without g-factor or adding MR noise. All models were trained for 50 epochs with the Sophia [6] optimizer and tested on a held-out test set of 560 samples. The test set contained 2D, 2D+T, and 3D images as well as different levels of MR noise. The MSE, L1, PSNR (max intensity 2048.0) and SSIM were.

*Cross field strength generalization, 0.55T MRI*. 8 healthy volunteers were scanned at a 0.55T MRI (MAGNETOM FreeMax, Siemens Healthcare) for R=2 and 3 retro-gated cine, CH4, CH2, CH3 and SAX stacks. To assess cross field strength generalization, the HRNet-imformer was applied to R=3 cine and ROIs were drawn in the LV and myocardium to measure SNR. The local point spread function [7] was measured on 452 points on the endo- and epi-cardiac boundaries to measure the resolution loss after the model. The cardiac function measurements (EF, ESV, EDV, MASS) were measured on model outputs and compared to R=2 measurements (without model), using a validated cine AI model [8].

*Cross imaging contrast and sequence generalization*. While training only used retro-gated cine 3T data, the model was applied to 1.5T CMR perfusion and LGE images (MAGNETOM Area, Siemens Healthcare).

*Cross anatomy generalization*. While all the training data was acquired on the heart, the model was applied to the three datasets to demonstrate cross anatomy performance: 1) T1 MPRAGE 3D neuro scan at 3T; 2) T2 TSE sagittal spine scan at 0.55T MRI; and 3) TSE sagittal knee scan at 0.55T MRI (prototype MAGNETOM Aera, Siemens, [9]). Note the latter two are also examples of cross field strength tests.

Results

Table 1 lists the model comparison results. The HRnet-imformer with the (TLG,TLG) block configuration gave the best performance, but quite comparable to the Unet-imformer. Among imformer models, adding spatial attention and noise augmentation improve performance over using only temporal attention or not using noise augmentation. Imaging transformers outperformed CNNs. Among the convolutional models, the 3D models performed better than the 2D models.

Figure 3 gives examples of 0.55T cines. Mean SNR gain for R=3 was 119-224% (90 percentiles) in the blood pool and 142-277% in the myocardium. The model was found to be locally linear (local linearity ratio[7] is 0.99±0.09) and LPSF was 1.04±0.11 for readout and phase, and 1.28±0.14 for temporal direction, indicating very slight spatial smoothing and the model learned to use temporal redundancy to recover signal from noise. Due to inferior quality, the cine AI model failed at 4 subjects for the raw R=3 images but was successful for all informer images. No significant differences were found between R=3 model outputs and R=2 images (paired t-test, P>0.15).

Figure 4 shows model generalization results on different imaging sequences (1.5T perfusion and LGE in 4a) and other anatomies (4b). While all training was performed on heart data from 3T scanners, the model generalized well.

Discussion

The model comparison results show the transformer models improved performance over convolutional models. This finding agrees with previous research in image classification [10] but has not been shown for MRI denoising. The proposed imaging transformer models also performed better than the conventional linear attention used in Swin3D. Further, the proposed imaging transformer blocks are flexible and able to process any combination of 2D, 2D+T or 3D data, make them promising model architectures for more general applications.

Conclusion

In this study, we a) propose a denoising training scheme consisting of SNR unit reconstruction and realistic noise augmentation, and b) propose novel imaging attention modules and shown their superior performance over CNN networks and conventional linear attention transformers for MRI denoising tasks. Together, these contributions result in strong generalization across field-strengths, scanners, imaging sequences, contrasts, and anatomies.

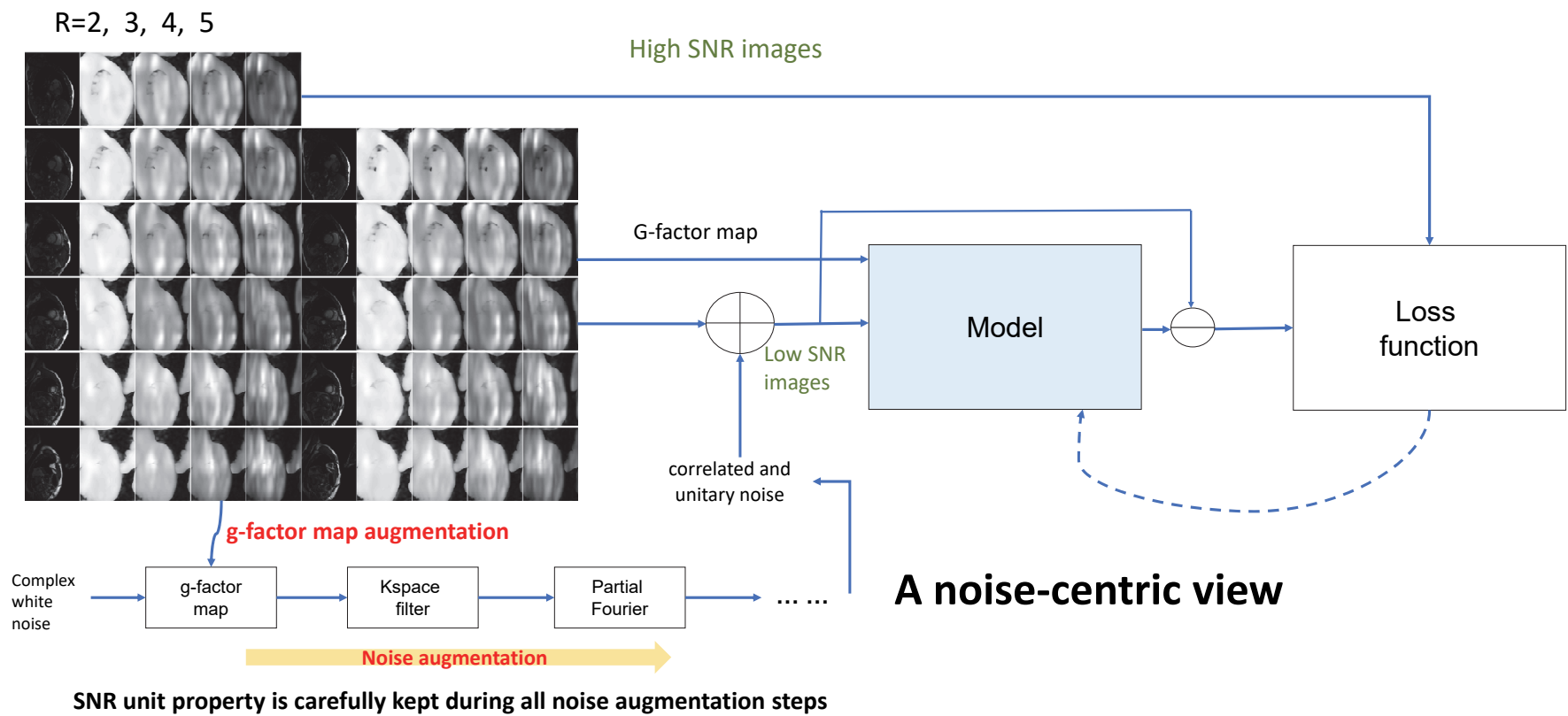

Figure 1. SNR unit training scheme. Training starts by computing the g-factor maps at different acceleration from the auto-calibration data of each training scans (g-factor map augmentation). The white, complex noise was transferred through different noise augmentation steps to follow the MR reconstruction pipeline. The SNRunit reconstruction was used here to maintain the noise std being unitary at every given step. The correlated and unitary noise was scaled up a g-factor map and added to the high-quality complex images. Different models can be plugged into this training scheme. Long-term skip connection was added to facilitate residual learning. The training was performed with two patch sizes (32x32 and 64x64, interleaved between steps). The loss was the sum of MSE, L1, perpendicular loss [ref] and PSNR loss. All models were trained with 50 epochs with the Sophia second-order optimizer [ref]. The model yields the lowest validation loss was picked for testing. 90% of the ~310K training samples were used for training and 10% were used for validation.

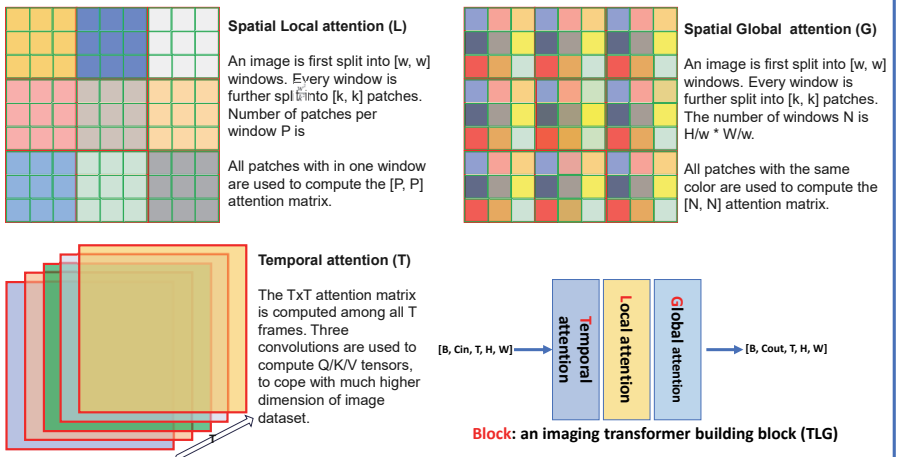

**(a) Imaging transformer (imformer) building blocks**

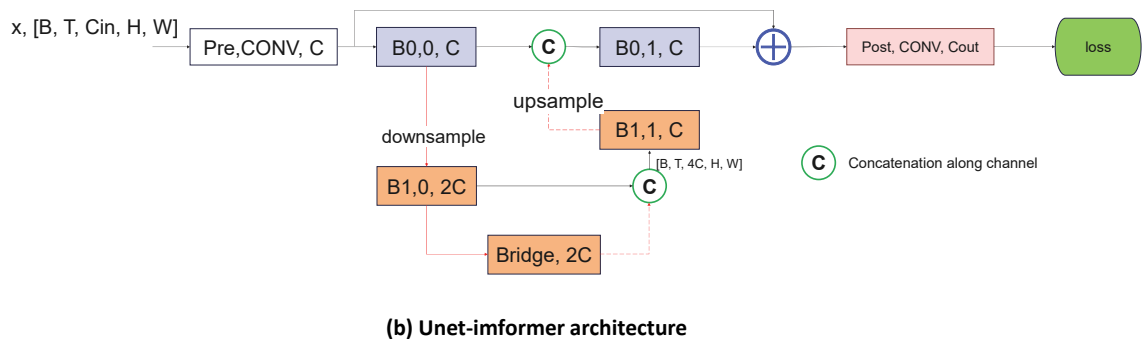

**(b) Unet-imformer architecture**

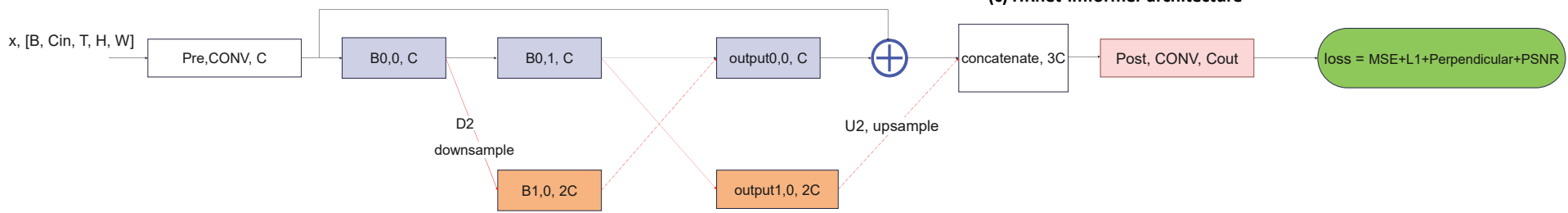

**(c) HRnet-imformer architecture**

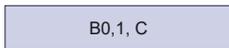

Block: a block is a basic unit of this model design. B0,1 means the second block on the resolution level 0 (original resolution). B1,0 means the first block at the resolution level 1 (downsample by 2 along H and W).

A block consists of several modules. They can be imaging attentions, e.g. TLG means a temporal attention, a local attention and a global attention. It can contain convolution layers as well, e.g., C2C2C2 means three concatenated components of [2D convolution, nonlinear activation, normalization]; C3C3C3 means concatenated three 3D convolutions.

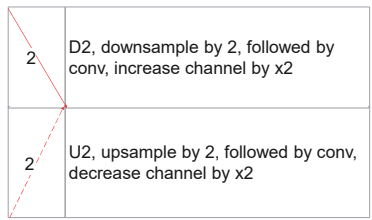

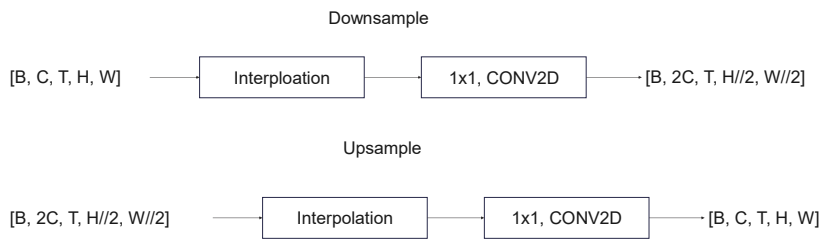

| HRnet, imformer | PSNR | SSIM | MSE | L1 | Unet, imformer | PSNR | SSIM | MSE | L1 |
|---|---|---|---|---|---|---|---|---|---|
| imformer, **Block-0**=TLG, **Block-1**=TLG (7.6M learnable parameters) | **73.094** | **0.8342** | **0.5618** | **0.6899** | TLG, TLG (8.2M learnable parameters) | **72.941** | **0.814** | **0.5812** | **0.7007** |
| TLG, TLG without g-factor augmentation | 67.216 | 0.2831 | 4.739 | 1.711 | TLG, TLG no g-factor maps augmentation | 68.237 | 0.7395 | 3.442 | 1.445 |
| TLG, TLG without noise augmentation | 63.376 | 0.3877 | 6.516 | 2.38 | TLG, TLG without noise augmentation | 66.722 | 0.6984 | 2.884 | 1.563 |
| TTT, TTT (7.6M) | 71.991 | 0.8124 | 1.444 | 0.8398 | TTT, TTT (8.2M) | 72.464 | 0.8085 | 0.8295 | 0.7653 |
| **HRnet, CNN** | PSNR | SSIM | MSE | L1 | **Unet, CNN** | PSNR | SSIM | MSE | L1 |
| C3C3C3,C3C3C3 (3.9M learning parameters) | 71.035 | 0.7921 | 2.057 | 0.9855 | C3C3C3,C3C3C3 (4.5M learnable parameters) | 71.054 | 0.7901 | 1.064 | 0.9004 |
| C3C3C3,C3C3C3 without g-factor augmentation | 68.29 | 0.2722 | 2.94 | 1.437 | C3C3C3,C3C3C3 without g-factor augmentation | 68.069 | 0.2686 | 3.591 | 1.49 |
| C3C3C3,C3C3C3 without noise augmentation | 66.338 | 0.3101 | 3.312 | 1.668 | C3C3C3,C3C3C3 without noise augmentation | 66.156 | 0.3119 | 3.402 | 1.694 |
| C2C2C2,C2C2C2 (1.5M learning parameters) | 66.269 | 0.6855 | 2.521 | 1.525 | C2C2C2,C2C2C2 (1.9M learnable parameters) | 66.594 | 0.6878 | 2.337 | 1.445 |
| **Conventional transformer** | PSNR | SSIM | MSE | L1 | | | | | |
| Swin 3D (15M) | 71.253 | 0.790 | 0.8695 | 0.8708 | | | | | |

Table 1. Model comparison results. Here **Block-0** and **Block-1** are the block structures for two resolution levels in HRnet and Unet setups. First, the HRnet-imformer gave slightly better results than the Unet-imformer. Second, the imformer models with both spatial and temporal attention outperforms the temporal only setup (TTT, TTT). Third, the imaging transformers outperform the convolution networks, with higher PSNR and SSIM, and lower MSE and L1. Fourth, the ablation tests were further performed to test the SNR unit training scheme. Inferior performance was found after turning off either g-factor map augmentation or MR noise augmentation, for both CNN and imformer models. Fifth, the HRnet-imformer outperformed the Swin 3D model which is a conventional linear attention transformer modified for imageries. Compared to the swin 3D, the imformer models had smaller number of learnable parameters. Finally, the 3D convolution models performed much better than 2D conv models.

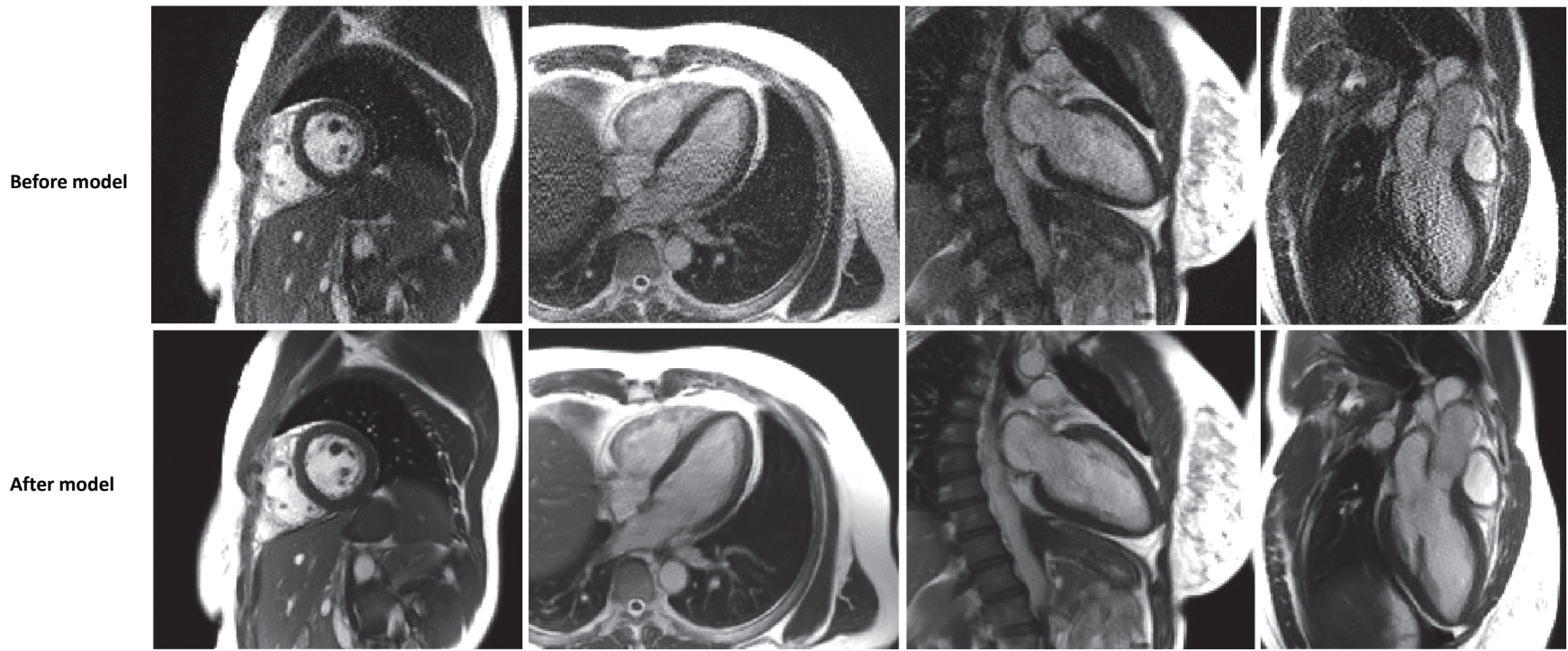

(a) 0.55T retro-gated cine results

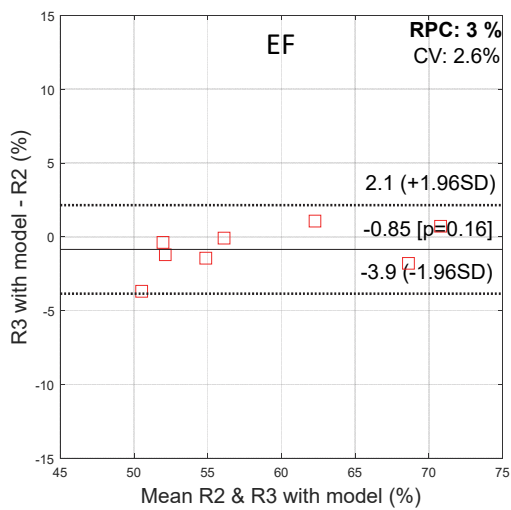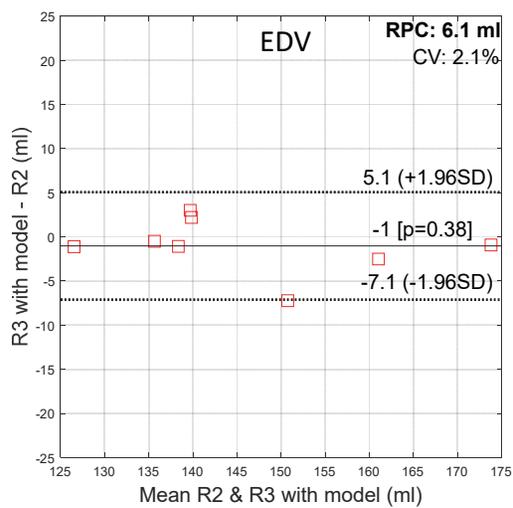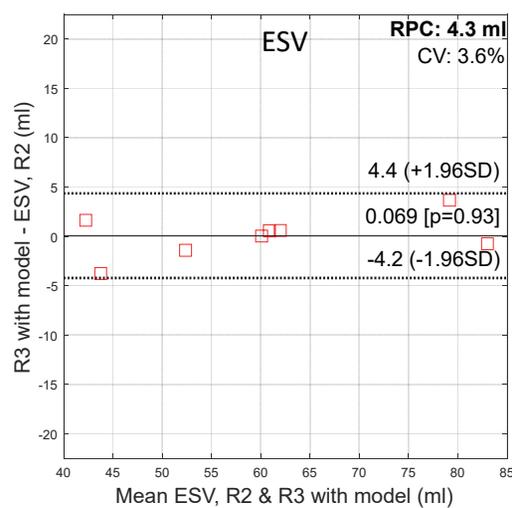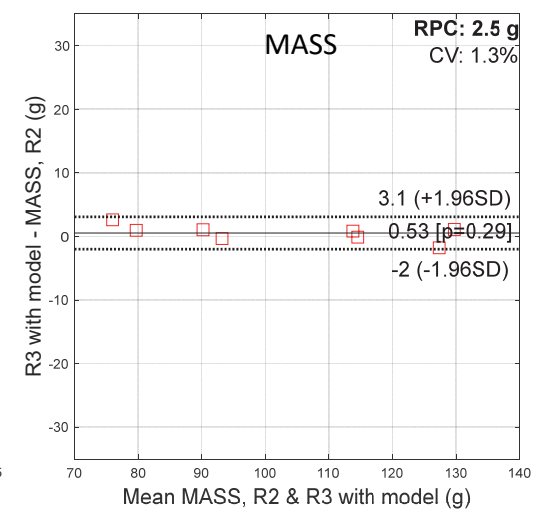

(b) Bland-Altman plots for cardiac function

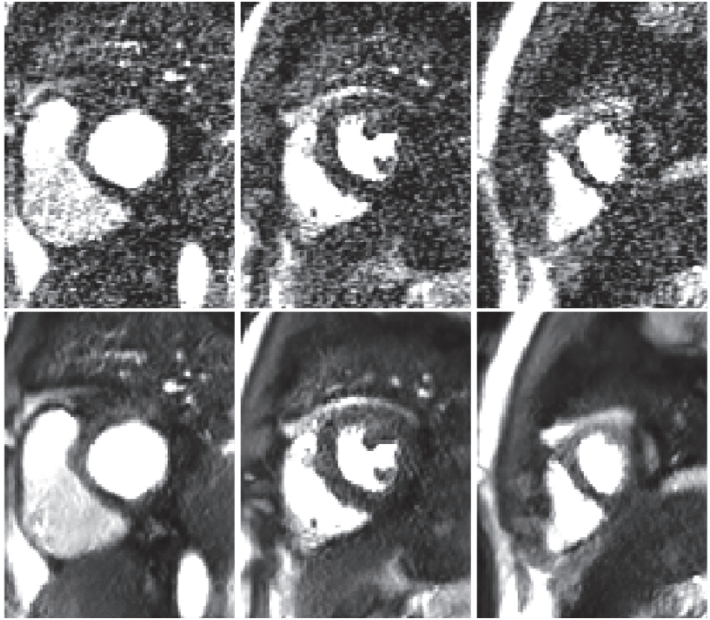

**(a) Cardiac perfusion imaging at 1.5T**

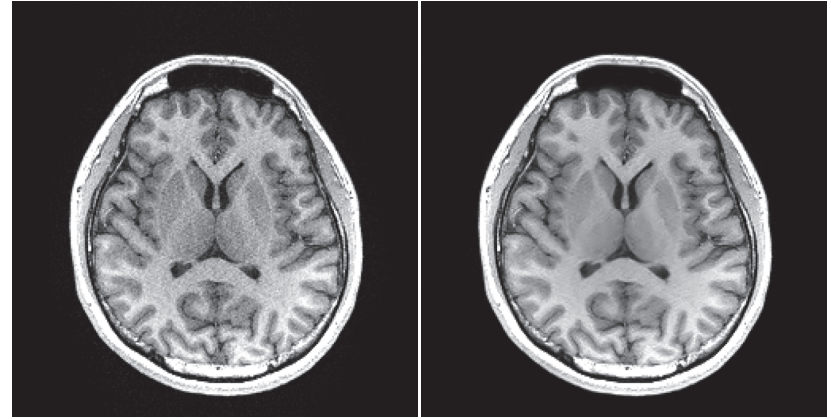

**(c) T1 MPRAGE neuro scan at 3T**

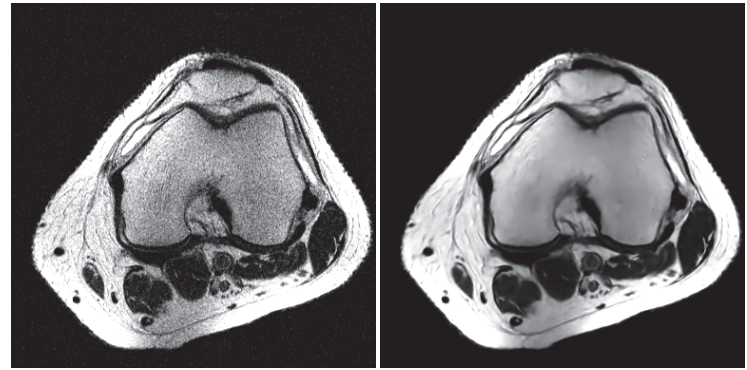

**(d) T2 TSE spine MRI at 0.55T**

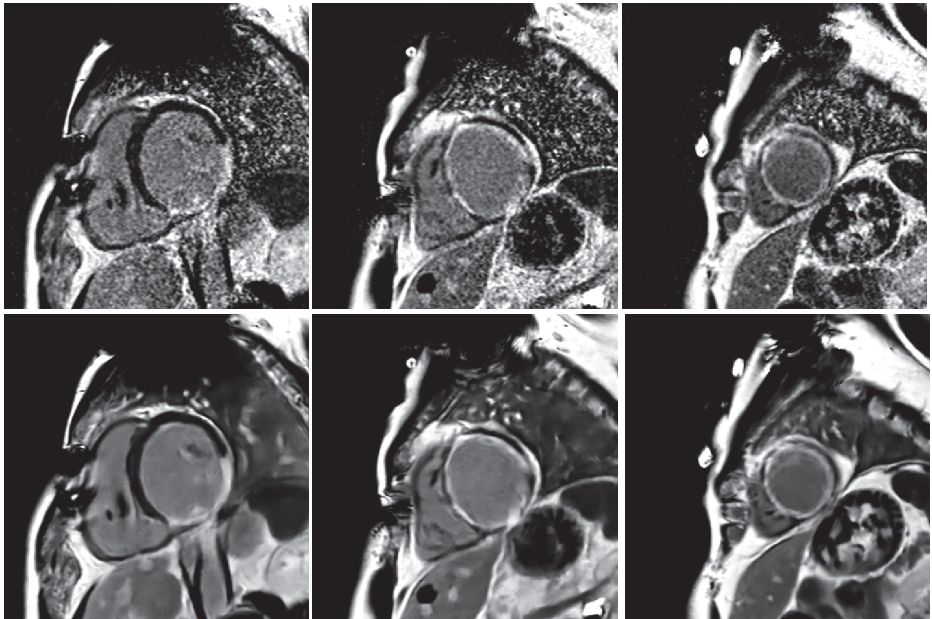

**(b) Free-breathing MOCO+AVE late Gd enhancement imaging at 1.5T**

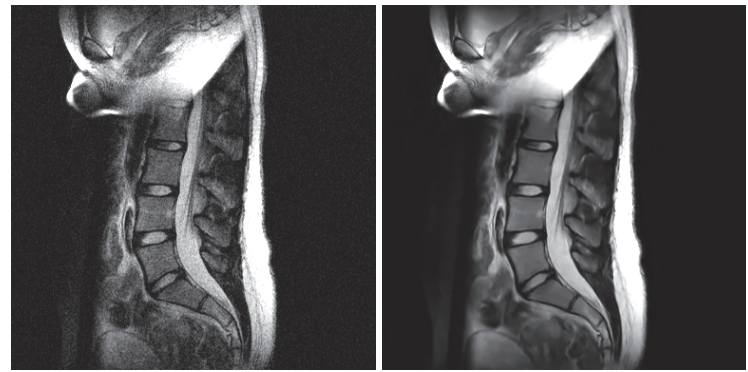

**(e) TSE knee MRI at 0.55T**